\documentclass[dvipdfmx]{article}

\usepackage{amssymb,amsfonts,amsmath,amsthm,stmaryrd}
\usepackage[nosort]{cite}
\usepackage{enumerate,float,indentfirst}
\usepackage[dvipdfmx]{graphicx}
\usepackage{tikz}
\usepackage{color}
\usepackage{bm}
\usepackage{simplewick}
\usepackage{here}
\usepackage{comment}

\catcode`\@=11
\@addtoreset{equation}{section}

\theoremstyle{definition}

\theoremstyle{remark}

\allowdisplaybreaks

\newcommand{\beq}{\begin{equation}}
\newcommand{\eeq}{\end{equation}}
\newcommand{\bea}{\begin{eqnarray}}
\newcommand{\eea}{\end{eqnarray}}
\newcommand{\mat}[1]{\begin{pmatrix} #1 \end{pmatrix}}
\def\nn{\nonumber}

\newcommand{\de}{\mathrm{d}}
\newcommand{\im}{\mathrm{i}}
\newcommand{\ex}{\mathrm{e}}

\definecolor{red}{rgb}{1,0,0}
\definecolor{orange}{rgb}{1,0.5,0}
\definecolor{violet}{rgb}{0.7,0,1}

\hoffset= - 1.0in         

\begin{document}

\begin{center}
\begin{small}
\hfill NITEP 144
\end{small}
\end{center}

\hfill May, 2023

\bigskip

\begin{center}{\Large{\bf  Construction of irregular conformal/W block \\
and flavor mass relations  of $\mathcal{N}=2$ SUSY gauge theory \\ 
 from the $\bm{A_{n-1}}$ quiver matrix model}}
\end{center}

\bigskip

\centerline{\large H. Itoyama$^{a, b,}$\footnote{e-mail: itoyama@omu.ac.jp}, 
T. Oota$^{a,b,}$\footnote{e-mail: toota@omu.ac.jp}, and 
R. Yoshioka$^{a,b,}$\footnote{e-mail: ryoshioka@omu.ac.jp} }

\vspace{1cm}

\noindent
{$^a$ \it Nambu Yoichiro Institute of Theoretical and Experimental Physics (NITEP),
			Osaka Metropolitan University (formerly Osaka City University)}\\
{$^b$ \it Osaka Central Advanced Mathematical Institute (OCAMI),
			Osaka Metropolitan University\\
\phantom{aaaaaaaa} 3-3-138, Sugimoto, Sumiyoshi-ku, Osaka, 558-8585, Japan}

\vspace{1cm}

\centerline{ABSTRACT}

\bigskip

  A sequence of massive scaling limits of the $\beta$-deformed $A_{n-1}$ quiver matrix model that keeps the size of the matrices finite and that corresponds to
   the $N_{f} =2n \rightarrow 2n-1, 2n-2$  limits on the number of flavors at 4d $su(n)$ ${\cal N} = 2$ SUSY gauge theory side is 
   carried out to provide us with the integral representation of $su(n)$ irregular conformal/W block. 
  The original paths are naturally deformed into those in the complex plane, 
   permitting us to convert into an $su(n)$ extension of the unitary matrix model of GWW type with a set of log potentials for   all species of eigenvalues.  
  Looking at the region in the parameter space that enjoys the maximal symmetry of the model,
  we derive a set of relations among the mass parameters which may serve as evidence 
  for the existence of the Argyres-Douglas critical hypersurface.

\bigskip

\bigskip

\section{Introduction}

Matrix models have played several important roles in the development of gauge theory, in particular, that of the ${\cal N}=2$ supersymmetric gauge theory (for reviews, see, for example, \cite{AG1991,GM9304,IY1507}). 
``Beta" deformed matrix ensembles derive the partition functions of such gauge theory \cite{Nekrasov2003} and assure the existence of the attendant curve\footnote{by this we mean the following: the Schwinger-Dyson eq. takes the form
of Virasoro constraints acting on the partition function. At the planar level, it implies
the regularity of the resolvent one-point function at $z=\infty$. This permits us to introduce a singularity-free object in ${\bf C}^2$, namely, a curve.}
 by the Schwinger-Dyson equation \cite{David1990,MM1990,IM1991}, embodying the integrability: 
in the case of $\beta =1$ (the case of free fermions), the partition function is by construction a tau function of a certain integrable hierarchy.

Matrix models also permit us to make exploit a variety of methods as statistical field theory (see, for example, textbooks \cite{ZJ}) and to probe critical phenomena and phase structure. While largely unexplored, combination of these two distinct disciplines appears to give us a fruitful avenue of thought.

The case of 2d Virasoro (or CFT) block whose integral representation is provided by one-matrix model of multi-log potential has been well-studied\cite{AGT0906,DV0909,IMO,EM0911,MMS0911,MMS1001,IO5,MMM1003,MS1004}.  
Much less studied are the  cases of $ W_n$ block (\cite{Wyllard0907,MM0908,IMO} 
and papers cited in therein) and for these cases, the series of multi- matrix models
of quiver type which obey the $ W_n$ constraints by construction \cite{MMM1991,KMMMP9208,Kostov9208} is available. 

Irregular limit \cite{Gaiotto0908,GT1203,MMM0909,IOYone1008,BMT1112,NR1112,NR1207,CR1312} of this series of models of the multi-log type is an interesting place to look at
 to explore new critical phenomena where an interplay between symmetry and phase structure  ala Landau is expected to play an essential role. 
 In this letter, we wish to report on a progress made in the construction of $A_{n-1}$ irregular block that is obeyed by the $W_n$/Virasoro constraints and  that is connected to susy gauge theory via 0d-4d connection.
We investigate broken symmetry structure of the model enforced by the automorphism of the Dynkin diagram and its possible connection to the critical hypersurface of the system, 
neither of which has been uncovered before despite the well-known notion of the singularities of the Argyres-Douglas type \cite{AD9505,APSW9511,KY9712,Xie1204}; 
these in principle can be  obtained from the Seiberg-Witten curve perse\footnote{QFT perse does not know the construction of the Seiberg-Witten curve itself, which requires the introduction of
ambient space ${\bf C}^2$.  The matrix model arising from the above-mentioned 2d-0d-4d connection realizes such curve as an eigenvalue distribution, which is eventually recognized as a discontinuity of multi-sheeted complex plane.  The matrix model not only provides
a construction of the curve but also permits us to compute various observables represented 
by (multi-point) resolvent correlators and to consider the problem of phases.}. 
We see that  imposing the maximal symmetry leads us to a set of flavor mass relations.

In the series of developments on the irregular limit of the $A_1$(one-matrix) case \cite{IOYanok1805,IOYanok1812,IOYanok2019,IOYanok1909,MOT1909,IYanok2103,Oota2112}, 
we have been led to the procedure of converting  the hermitean matrix model into the unitary matrix model in order to make such procedure well-defined. 
The upshot from the case of $N_f=2$ is the well-known GWW model \cite{GW1980,Wadia1980,Wadia1212,CIO1993} 
(see also \cite{Okuy1705,RT2007,Russo2010,ST2102} for recent discussion) augmented by the log potential. 
The Painlev\'e II equation with parameter has been derived in the double scaling limit \cite{IOYanok1805,IOYanok1812,IOYanok2019}.  
Let us note that the procedure is a conversion process that keeps the size of the matrix finite and is not just a simple universality argument \cite{Mizo0411}. 
Such  ``unitarization procedure" turns out to be responsible for making the above mentioned maximal symmetry manifest. 

In the next section, the $\beta$-deformed ADE quiver matrix model of three-log potential is briefly recalled.
In section three, we obtain $A_{n-1}$ irregular conformal/W block that corresponds to $su(n)$, $N_f=2n-1, 2n-2$, generalizing the work of \cite{IOYone1008,IOYanok1805,IOYanok1812,IOYanok2019,IOYanok1909,IYanok2103}.
In section four, after recasting the model into the unitary type, we observe the region in the  parameter space which enjoys the maximal symmetry enhancement of the model. 
In this region, we derive a set of interesting relations among the $2n-2$ flavor masses.
Furthermore, we can show that the Argyres-Douglas  hypersurface lies within this region of parameter space from Seiberg-Witten curve. 

\section{$\bm{\beta}$-deformed ADE quiver matrix model of three-log potential}
Let us briefly recall the $\beta$-deformed ADE quiver matrix model ($\beta = -b^2$): 
\begin{equation}\label{ADEmm}
Z \equiv \int \cdots \int \prod_{a=1}^r 
\left\{ \prod_{I=1}^{N_a} \de \lambda^{(a)}_I \right\}
\Bigl(\Delta_{\mathfrak{g}}(\lambda) \Bigr)^{-b^2}
\exp\left( - \frac{\im b}{g_s} \sum_{a=1}^r \sum_{I=1}^{N_a}
W_a(\lambda^{(a)}_I) \right), 
\end{equation}
where $W_a$ is a potential and 
\begin{equation}\label{Deltag}
\Delta_{\mathfrak{g}}(\lambda) = 
\prod_{a=1}^r \prod_{1 \leq I < J \leq N_a}
( \lambda_I^{(a)} - \lambda_J^{(a)})^2
\prod_{1 \leq a< b \leq r}
\prod_{I=1}^{N_a} \prod_{J=1}^{N_b}
( \lambda^{(a)}_I - \lambda^{(b)}_J )^{(\bm{\alpha}_a, \bm{\alpha}_b)}.
\end{equation}
Here, $\mathfrak{g}$ is a finite dimensional Lie algebra of ADE type with rank $r$, 
 $\mathfrak{h}$ the Cartan subalgebra of $\mathfrak{g}$ 
and $\mathfrak{h}^*$ its dual. 
We denote by $\bm{\alpha}_a \in \mathfrak{h}^*$ ($a = 1,2,\cdots,r$) the simple roots of $\mathfrak{g}$, $(\bullet,\bullet)$ being the inner product on $\mathfrak{h}^*$, 
which we normalize as $(\bm{\alpha}_a, \bm{\alpha}_a) = 2$. 

For an application to the Virasoro/W block and the 2d-4d connection through matrices, 
we choose the potential to be the three-Penner type. 
From now on, we restrict our attention to the Lie algebra of $A_{n-1}$ type.
Following  \cite{IMO}, let us introduce
\begin{equation} \label{multPenWa}
    W_a(z)
     =     \sum_{p = 1}^3 \left( \bm{\mu}_p, \bm{\alpha}_a \right)
           \log (q_p - z), 
\end{equation}
with $q_1 = 0$, $q_2=1$, $q_3=q$. 

Eq.\eqref{ADEmm} represents a generic 4 point block of $su(n)$ Toda theory. 
Such block has been originally introduced by $n-1$ independent free scalar fields  in the screening charge formalism, 
which is subsequently evaluated by the Wick contractions to give rise to eq.\eqref{ADEmm}.  
The $n-1$ species of screening charges with arbitrary numbers $N_1, N_2, \cdots$, and $N_{n-1}$  give us the momentum conservation:
\begin{equation}\label{mom-con}
\bm{\mu}_0 + \sum_{p=1}^3 \bm{\mu}_p + 
\sum_{a=1}^{n-1} {S}_a \bm{\alpha}_a = 0,
\end{equation}
where ${S}_a \equiv -\im b g_s N_a$. 
Following the recipe, we choose 
\begin{equation}
 \bm{\mu}_2 = {\mu}_{2,1} \bm{\Lambda}^1, ~~~
 \bm{\mu}_3 = {\mu}_{3,n-1} \bm{\Lambda}^{n-1}, 
\end{equation}
in accordance with the ``simple" puncture and 
$\bm{\mu}_0$, $\bm{\mu}_1$ to be a generic type written 
 in the bases of the fundamental weights $\bm{\Lambda}^a$:
\begin{equation}
 \bm{\mu}_p = \sum_a \mu_{p,a} \bm{\Lambda}^a. 
\end{equation}

\section{Limit to irregular block: $\bm{su(3)}$ and $\bm{N_f = 6 \to 5, 4}$ and the $\bm{su(n)}$ extension}
Let us first consider the sequence of limits of the model leading to the irregular conformal block, taking the case of $n=3$, $N_f=5, 4$. 

In \cite{IMO}, the 0d-4d dictionary has been given already at $n=3, N_f=6$  by comparing the Witten-Gaiotto curve with the matrix model curve:
\begin{equation}\label{dictionary:n=3}
\begin{split}
 (\bm{\mu}_0,\bm{\alpha}_1) = -m_1+m_2,~~~
 (\bm{\mu}_1,\bm{\alpha}_1) = m_4-m_5,~~~
 (\bm{\mu}_2,\bm{\alpha}_1) =  m_1+m_2+m_3,~~~
 (\bm{\mu}_3,\bm{\alpha}_1) = 0, \\ 
 (\bm{\mu}_0,\bm{\alpha}_2) = -m_2+m_3,~~~
  (\bm{\mu}_1,\bm{\alpha}_2) = m_5-m_6,~~~
  (\bm{\mu}_2,\bm{\alpha}_2) =  0,~~~
  (\bm{\mu}_3,\bm{\alpha}_2) = m_4+m_5+m_6. \\ 
\end{split}
\end{equation}
While this is at the planar level, 
it is adequate to tell that the appropriate decoupling limits from $N_f=6$ to 5 and 
subsequently to 4 are obtained by $m_6 \to \infty$ and subsequently by $m_1 \to \infty$. 
Other choices instead of $m_1$ and $m_6$, such as $m_2$ and $m_3$, 
involve $N_a \to \infty$ limit (see eq. \eqref{S:su(3)}), which spoils finiteness of a partial sum of masses according to the 0d-4d dictionary. 
This is why we do not consider the further limits $N_f = 3, 2,\cdots$ in this letter. 

Following the recipe of \cite{IO5} for the $su(2)$ case, 
 we specify  
 the integration contours for $su(3)$, $N_f=6$ case  to consist of $C_L = [0,q]$ and $C_R = [1,\infty]$. 
We adopt the following obvious generalization for 
$\de w_I^{a=1} {}_{I = 1,\cdots,N_1}$ and for $\de w_J^{a=2} {}_{J = 1,\cdots,N_2}$:
\begin{equation}
 C_{I, a}^{(6)} = \begin{cases}
 C_L ~~~\text{for } 1 \leq I \leq N_L^a,~ a=1,2, \\
 C_R ~~~\text{for } N_L^a+1 \leq I \leq N_a = N_L^a + N_R^a,~ a=1,2.
 \end{cases}
\end{equation}
The splitting of $N_a$ into $N_L^a$ and $N_R^a$ is related to the Coulomb moduli parameters  in the corresponding gauge theory and 
they remain fixed in the limits $m_1 \to \infty$ and $m_6 \to \infty$.

Thus our partition function for $su(n=3)$ $N_f = 6$ flavor case reads
\begin{align}\label{Z:n=3,N=6}
 Z_{A_{n-1}=A_2}^{(N_f=6)} = \text{(const)} &\int \cdots \int
 \prod_{a=1}^{n-1=2} \left\{ \prod_{I=1}^{N_a} \de w_I^{(a)} \right\} _{C_{I,a}^{(6)}}
 \Delta_{A_{n-1=2}}^{\beta}(w) \nn\\
 &\prod_{I=1}^{N_1} |w_I^{(1)}|^{\sqrt{\beta} (\bm{\mu}_1,\bm{\alpha}_1)}  
 |1-w_I^{(1)}|^{\sqrt{\beta} (\bm{\mu}_2,\bm{\alpha}_1)} 
 \prod_{J=1}^{N_2}  |w_J^{(2)}|^{\sqrt{\beta} (\bm{\mu}_1,\bm{\alpha}_2)} 
 |q-w_J^{(2)}|^{\sqrt{\beta} (\bm{\mu}_3,\bm{\alpha}_2)}. 
\end{align}

\begin{figure}[H]
\centering
\includegraphics[height=40mm]{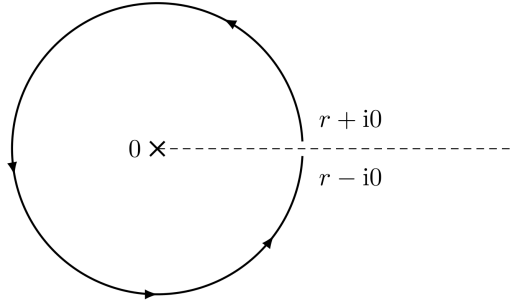}
\caption{The integration contour $C(r)$.}
\label{C(r)} 
\end{figure}

We now turn to the procedure of taking the $N_f=6 \to 5$ limit and subsequently the $N_f=5 \to 4$ limit. 
To handle infinities associated with these limits, it is necessary to deform the original paths on the real axis at $N_f=6$ to contours on the complex plane. 
In the case of the one-matrix model, we have devised and resolved this technical problem in \cite{IOYone1008,IOYanok1812}. 
Here in the case of multi-matrix model, we adopt the same procedure. 
For detail, we ask the reader interested to look at appendix F.2 and F.3 in \cite{IOYanok1812}. Our technique is, in fact, best summarized as the use of the following elementary integrations:
\begin{align}
 &\frac{1}{(\ex^{2\pi \im \alpha}-1)} \int_{C(q)} \de w w^{\alpha} \left( \sum_{n =0}^{\infty} c_n w^n\right)
 = \int_0^{q} \de w w^{\alpha} \left( \sum_{n =0}^{\infty} c_n w^n\right),~~~
 \text{for } \mathrm{Re}\,\alpha>0, \label{formula1}\\
 &\frac{1}{(1-\ex^{2\pi \im \alpha})} \int_{C(1)} \de w w^{\alpha} \left( \sum_{n =0}^{\infty} c_n w^{-n}\right)
 = \int_1^{\infty} \de w w^{\alpha} \left( \sum_{n =0}^{\infty} c_n w^{-n}\right),~~~
 \text{for } \mathrm{Re}\,\alpha<-1, \label{formula2}
\end{align}
where the contours $C(q)$ and $C(1)$ are depicted in Figure \ref{C(r)}.
The $N_f=6 \to 5$ limit corresponds to $\alpha = \sqrt{\beta} (\bm{\mu}_1,\bm{\alpha}_2) \to \infty $ and $ \left( \sum_{n =0}^{\infty} c_n w^n\right)$ being essentially the expansion of $(q-w)^b$ in \eqref{formula1}, and the infinity associated in the right hand side has been factorized in the left-hand side. This leads us to the expression \eqref{Z:n=3,N=5} in what follows. A similar procedure holds for the $N_f=5 \to 4$ limit that corresponds to $\alpha = \sqrt{\beta} (\bm{\mu}_2,\bm{\alpha}_1) \to \infty $ and $ \left( \sum_{n =0}^{\infty} c_n w^n\right)$ being essentially the expansion of $(1-1/w)^a$ in \eqref{formula2}. This leads us to the expression \eqref{Z:n=3,N=4} in what follows.

The $N_f = 6 \to 5$ limit is specified  by the following:
$m_6 \to \infty$, $q \to 0$ with $\Lambda_{5} = \lim q m_6$ (up to constant) kept finite.  
Let us introduce 
\begin{equation}
 q_{05} \equiv \lim q(\bm{\mu}_3,\bm{\alpha}_2), 
\end{equation}
so that $q_{05} = \Lambda_5 / g_s$ (up to constant). 

\begin{figure}[H]
\centering
\includegraphics[height=40mm]{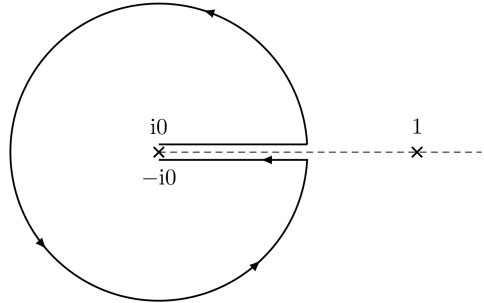}
\caption{The integration contour $C_0(r_0)$.}
\label{C0(r0)} 
\end{figure}

The resulting partition function for $N_f=5$ is 
\begin{align}\label{Z:n=3,N=5}
 Z_{A_2}^{(N_f=5)} = \text{(const)} &\int \!\!\cdots\!\! \int
 \prod_{a=1}^{2} \left\{ \prod_{I=1}^{N_a} \de w_I^{(a)} \right\}_{C_{I,a}^{(N_f =5)}}
 \Delta_{A_{2}}^{\beta}(w) \times  \nn\\
 &\times \prod_{I=1}^{N_1} (w_I^{(1)})^{\sqrt{\beta} (\bm{\mu}_1,\bm{\alpha}_1)}  
 (1-w_I^{(1)})^{\sqrt{\beta} (\bm{\mu}_2,\bm{\alpha}_1)} 
 \prod_{J=1}^{N_2}  (w_J^{(2)})^{\sqrt{\beta} (\bm{\mu}_1+\bm{\mu}_3,\bm{\alpha}_2)} 
 \exp \left( -\sqrt{\beta} \frac{q_{05}}{w_J^{(2)}} \right), 
\end{align}
the potential being
\begin{align}
 &W_1^{(N_f = 5)}(z) = (\bm{\mu}_1,\bm{\alpha}_1) \log z + (\bm{\mu}_2,\bm{\alpha}_2) \log(1-z), \\
 &W_2^{(N_f = 5)}(z) = (\bm{\mu}_1+\bm{\mu}_3,\bm{\alpha}_2) \log z - \frac{q_{05}}{z}. 
\end{align}
As for the complex contours $C_{I,a}^{(N_f =5)}$,  
\begin{equation}
 C_{I,a}^{(N_f=5)} = \left\{\!\!
 \begin{array}{ll}
  C_0(r_0) &\text{for } 1 \leq I \leq N_L^a,~~~a = 1,2,\\
  C(1) &\text{for } N_L^a \leq I \leq N_L^a + N_R^a,~~~a=1,2.
 \end{array}
\right.
\end{equation}
$C_0(r_0)$ is depicted in Figure \ref{C0(r0)}. The radius of the arc is $r_0$,
which is an arbitrary real number satisfying $0 < q < r_0 <1$. 
We have $w>0$, $1-w>0$ and $q-w>0$ for $w \in C_L$, and $w>0$, $1-w<0$ and $q-w<0$ for $w\in C_R$ in the integrand of \eqref{Z:n=3,N=6}. 
Since the sign of the quantities inside the modulus remain constant on the integration contour, 
 the absolute value have been dropped in \eqref{Z:n=3,N=5}.

Next, let us take the subsequent limit $N_f = 5 \to 4$: $m_1 \to \infty$, $q_{05} \to 0$ with  $\Lambda_4 \equiv \lim (m_1 \Lambda)^{1/2}$ (up to constant) kept finite. 

We introduce 
\begin{equation}
 q_{04}^2 \equiv \lim q_{05} (\bm{\mu}_2,\bm{\alpha}_1),
\end{equation}
so that $q_{04} = \Lambda_4/g_s$ up to some constant.

\begin{figure}[H]
 \centering
 \includegraphics[height=40mm]{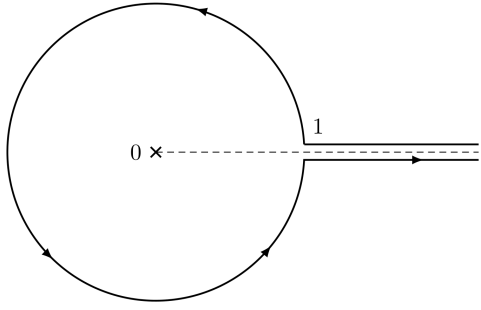}
 \caption{The integration contour $C_{\infty}$.}
 \label{C_infty}
 \end{figure}

The resulting partition function for $N_f=4$ is 
 \begin{align}
  Z_{A_2}^{(N_f=4)} = \text{(const)} &\int \!\!\cdots\!\! \int
  \prod_{a=1}^{2} \left\{ \prod_{I=1}^{N_a} \de w_I^{(a)} \right\}_{C_{I,a}^{(N_f=4)}}
  \Delta_{A_{2}}^{\beta}(w)  \times \nn\\
  &\times \prod_{I=1}^{N_1} (w_I^{(1)})^{\sqrt{\beta} (\bm{\mu}_1,\bm{\alpha}_1)}  
  \exp\left(-\sqrt{\beta} q_{04} w_I^{(1)} \right)
  \prod_{J=1}^{N_2}  (w_J^{(2)})^{\sqrt{\beta} (\bm{\mu}_1+\bm{\mu}_3,\bm{\alpha}_2)} 
  \exp \left( -\sqrt{\beta} \frac{q_{04}}{w_J^{(2)}} \right), 
  \label{Z:n=3,N=4}
 \end{align}
 the potential being
 \begin{align}
  &W_1^{(N_f = 4)}(z) = (\bm{\mu}_1,\bm{\alpha}_1) \log z -q_{04} z, \\
  &W_2^{(N_f = 4)}(z) = (\bm{\mu}_1+\bm{\mu}_3,\bm{\alpha}_2) \log z - \frac{q_{04}}{z}. 
 \end{align}
 As for the complex contours $C_{I,a}^{(N_f=4)}$, 
 \begin{equation}
  C_{I,a}^{(N_f=4)} = \left\{\!\! 
  \begin{array}{ll}
  C_0 = C_0(1) &\text{for } 1 \leq I \leq N_L^a,~~~a = 1,2,\\
  C_{\infty} &\text{for } N_L^a \leq I \leq N_L^a + N_R^a,~~~a=1,2,
  \end{array} \right.
 \end{equation}
 where $C_{\infty}$ is depicted in Figure \ref{C_infty}. 

It is completely straightforward to generalize the above discussion to $A_{n-1}$ quiver matrix model for $n \geq 4$. 
The starting point is the case $N_f =2n$:
\begin{align}
 Z_{A_{n-1}}^{(N_f=2n)} = (\text{const}) \int\!\!\cdots\!\!\int &\prod_{a=1}^{n-1} \left\{ \prod_{I_a=1}^{N_a} \de w_{I_a}^{(a)} \right\}_{C_{I,a}^{(N_f=2n)}} \left( \Delta_{A_{n-1}}(w) \right)^{\beta} \times \nn\\
&\times \left( \prod_{I_1=1}^{N_1} (w_{I_1}^{(1)})^{\sqrt{\beta} (\bm{\mu}_1,\bm{\alpha}_1)} (1-w_{I_1}^{(1)})^{\sqrt{\beta}(\bm{\mu}_2,\bm{\alpha}_1)} \right)  
\prod_{a=2}^{n-2} \prod_{I_a=1}^{N_a}  (w_{I_a}^{(a)})^{\sqrt{\beta} (\bm{\mu}_1,\bm{\alpha}_a)} \times \nn\\
&\times  \left( \prod_{I_{n-1}=1}^{N_{n-1}} (w_{I_{n-1}}^{(n-1)})^{\sqrt{\beta} (\bm{\mu}_1,\bm{\alpha}_{n-1})} (q-w_{I_{n-1}}^{(n-1)})^{\sqrt{\beta}(\bm{\mu}_3,\bm{\alpha}_{n-1})} \right). 
\end{align}
After the limit $N_f = 2n \to 2n-1$, we obtain 
\begin{align}
 Z_{A_{n-1}}^{(N_f=2n-1)} = (\text{const}) \int\!\!\cdots\!\!\int &\prod_{a=1}^{n-1} \left\{ \prod_{I_a=1}^{N_a} \de w_{I_a}^{(a)}  \right\}_{C_{I,a}^{(N_f = 2n-1)}} \left( \Delta_{A_{n-1}}(w) \right)^{\beta} \times \nn\\
&\times \left( \prod_{I_1=1}^{N_1} (w_{I_1}^{(1)})^{\sqrt{\beta} (\bm{\mu}_1,\bm{\alpha}_1)} (1-w_{I_1}^{(1)})^{\sqrt{\beta}(\bm{\mu}_2,\bm{\alpha}_1)} \right)
 \prod_{a=2}^{n-2} \prod_{I_a=1}^{N_a}  (w_{I_a}^{(a)})^{\sqrt{\beta} (\bm{\mu}_1,\bm{\alpha}_a)} \times \nn\\
&\times \left( \prod_{I_{n-1}=1}^{N_{n-1}} (w_{I_{n-1}}^{(n-1)})^{\sqrt{\beta} (\bm{\mu}_1 + \bm{\mu}_3 , \bm{\alpha}_{n-1})} \exp\left( -\sqrt{\beta} \frac{q_{05}}{w_{I_{n-1}}^{(n-1)}} \right) \right), 
\end{align}
where $W_1(z)$, $W_2(z)$, $a=2 \cdots, n-2$ are  the same as those at $N_f= 2n$ and 
\begin{align}
 &W_{n-1}(z) = (\bm{\mu}_1 + \bm{\mu}_3 , \bm{\alpha}_{n-1}) \log z - \frac{q_{05}}{z}, \\
 & q_{05} \equiv \lim q (\bm{\mu}_3 , \bm{\alpha}_{n-1}). 
\end{align}
After the limit $N_f = 2n-1 \to 2n-2$, we obtain 
\begin{align}
 Z_{A_{n-1}}^{(N_f=2n-2)} = (\text{const}) \int\!\!\cdots\!\!\int &\prod_{a=1}^{n-1} \left\{ \prod_{I_a=1}^{N_a} \de w_{I_a}^{(a)}  \right\}_{C_{I,a}^{(N_f = 2n-2)}} \left( \Delta_{A_{n-1}}(w) \right)^{\beta} \times \nn\\
&\times \left( \prod_{I_{1}=1}^{N_{1}} (w_{I_{1}}^{(1)})^{\sqrt{\beta} (\bm{\mu}_1 , \bm{\alpha}_{1})} \exp\left( -\sqrt{\beta} {q_{04}}{w_{I_{1}}^{(1)}} \right) \right)
 \prod_{a=2}^{n-2} \prod_{I_a=1}^{N_a}  (w_{I_a}^{(a)})^{\sqrt{\beta} (\bm{\mu}_1,\bm{\alpha}_a)} \times \nn\\
&\times \left( \prod_{I_{n-1}=1}^{N_{n-1}} (w_{I_{n-1}}^{(n-1)})^{\sqrt{\beta} (\bm{\mu}_1 + \bm{\mu}_3 , \bm{\alpha}_{n-1})} \exp\left( -\sqrt{\beta} \frac{q_{04}}{w_{I_{n-1}}^{(n-1)}} \right) \right),  
\end{align}
where $q_{04}^2 = \lim q_{05} (\bm{\mu}_2 , \bm{\alpha}_1)$. 
The potentials are 
\begin{equation}
\begin{split}
 &W_1(w^{(1)}) = (\bm{\mu}_1 , \bm{\alpha}_1) \log w^{(1)} - q_{04} w^{(1)}, \\
 &W_a(w^{(a)}) = (\bm{\mu}_1 , \bm{\alpha}_a) \log w^{(a)},~~~~~~~~~a = 2, \cdots, n-2, \\
 &W_{n-1}(w^{(n-1)}) = (\bm{\mu}_1 + \bm{\mu}_3 , \bm{\alpha}_{n-1}) \log w^{(n-1)} - \frac{q_{04}}{ w^{(n-1)}}.
\end{split}
\end{equation}
In this form, the (broken) symmetry of the model based on the even parity of the potential and the automorphism of the Dynkin diagram is still not manifest. 

The 0d-4d dictionary worked out in \cite{IMO} for the $A_{n-1}$ case reads 
\begin{equation}\label{dictionary}
\begin{split}
 &\bm{\mu}_0 = \sum_{a=1}^{n-1} (-m_a + m_{a+1}) \bm{\Lambda}^a,~~~
 \bm{\mu}_1 = \sum_{a=1}^{n-1} (m_{n+a} - m_{n+a+1}) \bm{\Lambda}^a,\\
& \bm{\mu}_2 = \left( \sum_{i =1}^n m_i \right) \bm{\Lambda}^1,~~~
 \bm{\mu}_3 = \left( \sum_{i =1}^n m_{i+n} \right) \bm{\Lambda}^{n-1}. 
\end{split}
\end{equation}
Further limit to
$ N_f=2n-3$ and the lower ones are not available at the current approach by the same reason as was mentioned before at \eqref{dictionary:n=3}.

\section{The coefficients after ``unitarization"}
As was stated in the introduction, we would like to recast the above $A_{n-1}$ quiver matrix model 
 for the irregular Virasoro/W block into the unitary form:
\begin{align}\label{Z:n-1}
&Z_{A_{n-1},U}^{N_f=2n-2} \equiv \int\!\!\cdots\!\!\int \prod_{a=1}^{n-1} \left\{ \prod_{I_a=1}^{N_a} \frac{\de w_{I_a}^{(a)}}{w_{I_a}} \right\}_{C_{I,a}^{N_f=2n-2}} \left( \Delta_{A_{n-1}}(w) \right)^{\frac{\beta}{2}} \left( \Delta_{A_{n-1}}\left(\frac{1}{w}\right) \right)^{\frac{\beta}{2}}  \times \nn\\
&\times \prod_{I_1=1}^{N_1} (w_{I_1}^{(1)})^{\sqrt{\beta} c_1} \exp\left( -\sqrt{\beta} q_{04} w_{I_1}^{(1)} \right) \prod_{a=2}^{n-2} \prod_{I_a=1}^{N_a} (w_{I_a}^{(a)})^{\sqrt{\beta} c_a}  
\prod_{I_{n-1}=1}^{N_{n-1}} (w_{I_n-1}^{(n-1)})^{\sqrt{\beta} c_{n-1}} \exp\left( -\sqrt{\beta}\frac{ q_{04}}{ w_{I_{n-1}}^{(n-1)}} \right). 
\end{align}
The potentials have the following form: 
\begin{equation} \label{potential}
\begin{split}
&W_{U,1} (w^{(1)}) = c_1 \log w^{(1)} - q_{04} w^{(1)}, \\
&W_{U,a} (w^{(a)}) = c_a \log w^{(a)}, ~~~~~~~~~~a = 2, \cdots, n-2,\\
&W_{U,n-1} (w^{(n-1)}) = c_{n-1} \log w^{(n-1)} - \frac{q_{04}}{ w^{(n-1)}}. 
\end{split}
\end{equation}

Let us turn to the evaluation of $c_a$.  
Recasting the partition function in the form \eqref{Z:n-1} brings the extra contribution to the potential, 
 from which we deduce the coefficients $c_a$. 
For definiteness, let us again first consider the case of $su(3)$. 
Combining \eqref{mom-con} with \eqref{dictionary:n=3}, 
 we can readily obtain 
\begin{equation}\label{S:su(3)}
 - \im  b g_s N_1 ={S}_1 = -m_2-m_3-m_4,~~~~~
 - \im  b g_s N_2 ={S}_2 = -m_3-m_4-m_5.
\end{equation}
Hence, 
\begin{align}
 c_1 &= (\bm{\mu}_1,\bm{\alpha}_1) + {S}_1 + \frac{(\bm{\alpha}_1,\bm{\alpha}_2)}{2} {S}_2 \nn\\
 &= (m_4-m_5) - (m_2+m_3+m_4) +\frac{1}{2} (m_3+m_4+m_5) \nn\\
 &= -\left(m_2-m_4 + \frac{1}{2}(m_3+m_4+m_5)\right), \\
 c_2 
 &= (\bm{\mu}_1+\bm{\mu}_3, \bm{\alpha}_2) + \frac{(\bm{\alpha}_1,\bm{\alpha}_2)}{2} {S}_1  + {S}_2 \nn\\
 &= (m_{3+2} - m_{3+2+1}) + \sum_{i =1}^3 m_{i+3}  + \frac{1}{2}(m_2+m_3+m_4) -(m_3+m_4+m_5) \nn\\
 &= m_5-m_3 + \frac{1}{2}(m_2+m_3+m_4).
\end{align}
Let us define $\check{c}_a $ by replacing  $\bm{\mu}_1 + \bm{\mu}_3$ in $c_a$ with  $\bm{\mu}_0 + \bm{\mu}_2$. 
Note that $(\bm{\mu}_1 + \bm{\mu}_3,\bm{\alpha}_1) = (\bm{\mu}_1,\bm{\alpha}_1)$. 
We observe that
\begin{align}
 \check{c}_1 &\equiv (\bm{\mu}_0+\bm{\mu}_2,\bm{\alpha}_1) + {S}_1 +
 \frac{(\bm{\alpha}_1,\bm{\alpha}_2)}{2} {S}_2 \nn\\
 &= -m_1+m_2+\sum_{i =1}^3 m_i -(m_2+m_3+m_4) + \frac{1}{2}(m_3+m_4+m_5) \nn\\
 &= m_2-m_4 + \frac{1}{2} (m_3+m_4+m_5) = -c_1. 
\end{align}
This is actually a relation valid for arbitrary $n$:
\begin{align}
 \check{c}_1 + c_1 
 & = (\bm{\mu}_0 + \bm{\mu}_2 + \bm{{\mu}_1, \bm{\alpha}_1})
 +2 {S}_1 + {(\bm{\alpha}_1,\bm{\alpha}_2)} {S}_2 \nn\\
 &\underset{\eqref{mom-con}}{=} -( \bm{\mu}_3 + {S}_1 \bm{\alpha}_1 + {S}_2 \bm{\alpha}_2, \bm{\alpha}_1)
 +2 {S}_1 + {(\bm{\alpha}_1,\bm{\alpha}_2)} {S}_2 \nn\\
 &\underset{\eqref{dictionary:n=3}}{=} 0.
\end{align}
So, by the ``unitarization" procedure, 
we have restored the left-right symmetry of the original irregular 4 point block, namely,
the symmetry under  $\bm{\mu}_0 + \bm{\mu}_2   \leftrightarrow \bm{\mu}_1 + \bm{\mu}_3$
 which has become invisible by the $SL(2,\mathbb{C})$ fixing. 
Let us note that, in the  known case $su(2)$, $N_1=-m_2-m_3$ and 
\begin{align}
 &\check{c}_1 = -m_1+m_2 + \sum_{i =1}^2 m_i + {S}_1 = m_2-m_3, \\
 &c_{n-1} = c_1 = m_3-m_4 + \sum_{i =1}^2 m_{i+2} + {S}_1 = m_3-m_2.
\end{align}

One can proceed further to work out the coefficients $c_1 = -\check{c}_1, c_2, \cdots, c_{n-1}$ for the general case $su(n)$ from $N_1, \cdots,N_{n-1}$ and the dictionary \eqref{dictionary}. 
Note that eq.\eqref{mom-con} is inverted to give 
\begin{equation}
\mat{{S}_1 \\ {S}_2 \\ \vdots \\ {S}_{n-1}}=  -\mathcal{C}^{-1} 
\mat{(\bm{\alpha}_1,\sum_i \hat{\bm{\mu}}_i) \\ (\bm{\alpha}_2,\sum_i \hat{\bm{\mu}}_i) \\
\vdots \\  (\bm{\alpha}_{n-1}, \sum_i \hat{\bm{\mu}}_i)} , 
\end{equation}
where 
\begin{equation}
\mathcal{C}^{-1}_{a,a'} = (\bm{\Lambda}_a, \bm{\Lambda}_{a'})
 = \frac{1}{n} \min (a,a') \{ n - \max (a,a') \},
\end{equation}
is the inverse of the Cartan matrix. 

For $n=4$, we obtain 
\begin{equation}
 {S}_1 = -(m_2+m_3+m_4+m_5), ~~~
 {S}_2 = -(m_3+m_4+m_5+m_6), ~~~
 {S}_3 = -(m_4+m_5+m_6+m_7),
\end{equation}
\begin{align}
 \check{c}_1 &= m_2-m_5 + \frac{1}{2}(m_3+m_4+m_5+m_6) \nn\\
 &= m_2 + \frac{1}{2} (m_3+m_4-m_5+m_6), \\
 \check{c}_2&\equiv (\bm{\mu}_0 + \bm{\mu}_2, \bm{\alpha}_2) -\frac{1}{2} {S}_1 + {S}_2 -\frac{1}{2} {S}_3\nn\\
 &= -m_2 -m_6 + \frac{1}{2} (m_2+m_3) + \frac{1}{2} (m_6+m_7) \nn\\
 &= \frac{1}{2}(-m_2+m_7) + \frac{1}{2}(m_3-m_6), \\
 c_3&= m_7-m_4 + \frac{1}{2}(m_3+m_4+m_5+m_6) \nn\\
  &= m_7 + \frac{1}{2} (m_3-m_4+m_5+m_6). 
\end{align}
For general $n$, 
\begin{align}
 c_1 &= (\bm{\mu}_1,\bm{\alpha}_1) + {S}_1 + \frac{(\bm{\alpha}_1 , \bm{\alpha}_2)}{2} {S}_2, \\
 &\vdots \nn\\
 c_a &= (\bm{\mu}_1 + \bm{\mu}_3 , \bm{\alpha}_a) + \frac{(\bm{\alpha}_a , \bm{\alpha}_{a-1})}{2} {S}_{a-1} + {S}_a + \frac{(\bm{\alpha}_a , \bm{\alpha}_{a+1})}{2} {S}_{a+1}, \nn\\
 &~~~~~\text{with } (\bm{\mu}_3 , \bm{\alpha}_a) =0,~~~ a=2,\cdots,n-2, \\
 &\vdots \nn \\
 c_{n-1} &= (\bm{\mu}_1 + \bm{\mu}_3 , \bm{\alpha}_{n-1}) + \frac{(\bm{\alpha}_{n-1} , \bm{\alpha}_{n-2})}{2} {S}_{n-2} +{S}_{n-1}. 
\end{align}
Let 
\begin{align}
 \check{c}_1 &\equiv (\bm{\mu}_0 + \bm{\mu}_2 , \bm{\alpha}_1) + {S}_1 + \frac{(\bm{\alpha}_1 , \bm{\alpha}_2)}{2} {S}_2, \\
 &\vdots \nn\\ 
 \check{c}_a &\equiv (\bm{\mu}_0 + \bm{\mu}_2 , \bm{\alpha}_a) +  \frac{(\bm{\alpha}_a , \bm{\alpha}_{a-1})}{2} {S}_{a-1} + {S}_a + \frac{(\bm{\alpha}_a , \bm{\alpha}_{a+1})}{2} {S}_{a+1},
 ,~~~ a=2,\cdots,n-2, \\
 &\vdots \nn\\
 \check{c}_{n-1} &\equiv (\bm{\mu}_0 + \bm{\mu}_2 , \bm{\alpha}_{n-1}) + \frac{(\bm{\alpha}_{n-1} , \bm{\alpha}_{n-2})}{2} {S}_{n-2} +{S}_{n-1}.
\end{align}
We have checked 
\begin{equation}
 c_a = - \check{c}_a, ~~~~~ \text{for } a= 1, \cdots, n-1,
\end{equation}
namely,  that $c_a$ has a dual expression $-\check{c}_a$. 
We obtain 
\begin{align}
 &{S}_1 = -(m_2+\cdots+m_{n+1}), ~~~
 {S}_2 = -(m_3+\cdots+m_{n+2}), ~~~ 
 \cdots, ~~~\nn\\
 &~~~ ~~~ {S}_{n-2} = -(m_{n-1}+\cdots+m_{2n-2}), ~~~
 {S}_{n-1} = -(m_{n}+\cdots+m_{2n-1}),
\end{align}
and 
\begin{align}
\check{c}_1 &= m_2 - m_{n+1} + \frac{1}{2} \sum_{i =1}^n m_{2+i}, \\
\vdots& \nn\\ 
\check{c}_a &= (-m_a + m_{a+1}) + \frac{1}{2}(m_a - m_{a+1}) -\frac{1}{2} (m_{a+n} - m_{a+n+1}) \nn\\
 &= -\frac{1}{2} (m_a - m_{a+1}) - \frac{1}{2} (m_{a+n} - m_{a+n+1} ),~~~ a=2,\cdots,n-2, \\ 
\vdots& \nn\\
\check{c}_{n-1} &= m_{2n-1} - m_n + \frac{1}{2} \sum_{j=1}^n m_{2n-1-j}. 
\end{align}

\section{Constraints of maximal symmetry}
We would now like to impose the maximal symmetry which the system can have in some region of the parameter space and to derive a set of relations among the mass parameters $m_2, m_3, \cdots$ and $m_{2n-1}$.

Let us recall that, in $n=2$, the case of $su(2)$, $c_1 \log w^{(1)}$ in eq.\eqref{potential} is the term that breaks even parity of the potential $W_{U,1}(w^{(1)})$ under $w^{(1)} = 1/ w^{(1)\prime}$.
Demanding the invariance under this inversion, we obtain
\begin{equation}
 c_1 = -\check{c}_1 = 0, 
\end{equation}
which means $m_2 = m_3$, namely, the Argyres-Douglas  point. 

For $n=3$, the case of $su(3)$, let us demand the symmetry under a generalized inversion that reflects the combination of even parity and folding of the Dynkin diagram:
\begin{equation}
 w^{(1)} = \frac{1}{w^{(2)\prime}},~~~~~
 w^{(2)} = \frac{1}{w^{(1)\prime}}. 
\end{equation}
This gives us 
\begin{equation}
 {S}_1 = {S}_2 ~~~~~\text{and}~~~~~ \check{c}_1 = c_2,
\end{equation}
which imply
\begin{equation}
 m_2 = m_5,~~~~~ m_3=m_4. 
\end{equation}

Likewise, for $n=4$, the case of  $su(4)$, demand the  invariance under
\begin{equation}
 w^{(1)} = \frac{1}{w^{(3)\prime}},~~~~~
  w^{(3)} = \frac{1}{w^{(1)\prime}},~~~~~
  w^{(2)} = \frac{1}{w^{(2)\prime}}. 
\end{equation} 
This gives us 
\begin{equation}
 {S}_1 = {S}_3,~~~~~
 \check{c}_1 = c_3,~~~~~
 \check{c}_2 = c_2 = 0. 
\end{equation}
These respectively imply 
\begin{equation}
 m_2-m_7 = -(m_3-m_6),~~~~~
 m_2-m_7 = -(m_4-m_5),~~~~~
 m_2-m_7 = m_3-m_6.
\end{equation}
We conclude 
\begin{equation}
 m_2-m_7 = m_3-m_6 = m_4-m_7 = 0. 
\end{equation}

This interesting pattern persists at $n=5$. 
Demanding the invariance, we obtain
\begin{equation}
 {S}_1 = {S}_4,~~~~~
 {S}_2 = {S}_3,~~~~~
 \check{c}_1 = c_4,~~~~~
 \check{c}_2 = c_3.
\end{equation}
These respectively imply 
\begin{equation}
\begin{split}
& (m_2-m_4) + (m_3-m_8) + (m_4-m_7) = 0,\\
&m_3 - m_8 = 0,~~~~~
 m_2-m_9 = -(m_5-m_6), ~~~~~
 m_2-m_9 = m_3-m_8. 
\end{split}
\end{equation}
We conclude 
\begin{equation}
 m_2-m_9 = m_3-m_8 = m_4-m_7 = m_5-m_6 = 0. 
\end{equation}

At general $n$, there are $2n-2$ mass parameters, and the conditions to be imposed are the invariance under $w^{(a)} = 1/w^{(n-a)\prime}, a=1,\cdots,n-1$.
For $n$ odd, there are $\frac{n-1}{2}$ conditions of ${S}_a = {S}_{n-a}$, $a=1, \cdots, \frac{n-1}{2}$, and $\frac{n-1}{2}$ conditions of $\check{c}_a = c_{n-a}$, $a=1,\cdots,\frac{n-1}{2}$.
For $n$ even, there are $\frac{n}{2}-1$  conditions of ${S}_a ={S}_{n-a}$, $a = 1, \cdots, \frac{n}{2}-1$, 
and $\frac{n}{2}$ conditions of $\check{c}_a = c_{n-a}$, $a = 1, \cdots, \frac{n}{2} -1$ 
and $\check{c}_{n/2} = c_{n/2} = 0$. 
In both cases, the number of net free parameters is $n-1$.
Doing the same analysis as that in $n=4$ for $n$ even, and that in $n=5$ for $n$ odd, we conclude
\begin{equation}\label{MassRelation}
 m_i = m_{2n+1-i},~~~~~i=2,\cdots, n.
\end{equation}

It is interesting to see that, after imposing the maximal symmetry of the system, 
we still do not reach the point of all masses being equal, 
 which is sometimes assumed in the analysis of the Seiberg-Witten curve in general. 
As the matrix model curve of this model and the corresponding Witten-Gaiotto curve are known to be isomorphic \cite{IMO}, the same conclusion should be drawn at the level of the Seiberg-Witten curve perse.

For example,  
combining the Seiberg-Witten curve for $su(n)$, $N_f=2n-2$ case presented in \cite{HO9505} with \eqref{MassRelation}, we can show (for detail, see \cite{IOYoshi}) that, under the appropriate conditions on the parameters,  the Seiberg-Witten curve can be written as
\begin{equation}\label{curve:degenerate}
 y^2 = x^{n+1} \left\{ 
  x^{n-1} \pm 2\Lambda \sum_{a=0}^{n-2} e_a x^{n-2-a} 
 \right\}.
\end{equation}
Here $\Lambda$ is the scale parameter and
 $e_a= e_a(\tilde{m}_2,\cdots, \tilde{m}_{n-1})$  is the elementary symmetric polynomial in $\tilde{m}_{i} = m_{i} - m_{n}$. 
The curve is maximally degenerate with multiplicity $n+1$. 
The original Seiberg-Witten curve has $2n-1$ parameters, which consist of $n-1$ moduli parameters, $n-1$ mass parameters and the scale parameter.  
In order to derive \eqref{curve:degenerate}, 
we need $n$ conditions and these define  Argyres-Douglas  critical  hypersurface in the parameter space.

\section*{Acknowledgments}
We thank Katsuya Yano for the series of collaborations on this subject. 
The work of HI is supported in part by JSPS KAKENHI Grant Number 19K03828 and by
the Osaka City University (OCU) Strategic Research Grant 2021 for priority area (OMU-SRPP2022\_SU04).





\end{document}